# Identifying Coordination Problems in Software Development: Finding Mismatches between Software and Project Team Structures

Chintan Amrit, Jos van Hillegersberg, and Kuldeep Kumar

**Abstract**— Today's dynamic and iterative development environment brings significant challenges for software project management. In distributed project settings, "management by walking around" is no longer an option and project managers may miss out on key project insights. The TESNA (TEchnical Social Network Analysis) method and tool aims to provide project managers both a method and a tool for gaining insights and taking corrective action. TESNA achieves this by analysing a project's evolving social and technical network structures using data from multiple sources, including CVS, email and chat repositories. Using pattern theory, TESNA helps to identify areas where the current state of the project's social and technical networks conflicts with what patterns suggest. We refer to such a conflict as a Socio-Technical Structure Clash (STSC). In this paper we report on our experience of using TESNA to identify STSCs in a corporate environment through the mining of software repositories. We find multiple instances of three STSCs (Conway's Law, Code Ownership and Project Coordination) in many of the on-going development projects, thereby validating the method and tool that we have developed.

――――――――――― ◆ ―――――――――――

## INTRODUCTION

While there is no single cause for the problems in Software Development, a major factor is the problem of co-ordinating activities while developing software systems [1]. The inherent complexity, conformity, changeability and invisibility of software make coordination particularly hard [2], in both large and small collocated projects [3, 4]. A particular kind of coordination problem that is pervasive in the software development environment occurs due to a misalignment of the social and technical structures of an organization. We call such a misalignment a Socio-Technical[1] Structure Clash (STSC).

In response to some of these coordination problems, researchers have developed detailed Organizational and Process patterns for describing the preferred relationships between the team communication structure (the social network) and the technical software architecture [7, 8]. As the term Process Pattern [7] is also used in business process management and workflow, we prefer to use the term Socio-Technical Pattern to refer to those patterns involving coordination problems related to both the social and technical aspects. Socio-Technical Patterns are a subset of the Organizational and Process patterns described by Coplien and Harrison [7]. A STSC is the specific coordination problem addressed by a Socio-Technical Pattern. In other words, an STSC occurs if the social network of the software development team does not match the Socio-Technical dependencies in the software architecture [3, 9].

As Socio-Technical Patterns are based on a wide variety of knowledge and experience, they are potentially very useful for a project manager in planning and monitoring complex development projects. Though many of these patterns have been around for a while in the field of Software Engineering [8], in practice these patterns are difficult to implement and monitor. The reason behind this is that it is difficult to find coordination problems in order to apply the solutions provided by the Socio-Technical Patterns, as purely manual techniques are labour intensive. Especially within dynamic, iterative and large scale projects the use of Socio-Technical Patterns is challenging. Social and technical data can constantly change making it hard for the project manager to keep track of different STSCs. Even in a small company employing less than 20 developers the social network and the rela-tion to the software tasks can get quite complicated for the software manager to track [3]. In larger scale projects the volume of social and technical data makes it even more challenging to identify STSCs. Most of these large scale projects have large number of people, and hence the size and number of social networks along with the size of the technical networks hinder effective identification of STSCs. The continuous and early detection of STSCs can help project managers in monitoring the software development process and enable managers to take corrective measures whenever STSCs occur [10].

---

[1] Here we use the Socio-Technical to refer to the different degrees of Social and Technical nature of project work. The term is directly based on the Socio-Technical Interaction Network or STINs by [5] R. Kling, G. W. McKim, and A. Kin, "A Bit More to It: Scholarly Communication Forums as Socio-technical Interaction Networks," *JASTIS*, vol. 54, no. 1, pp. 47-67, 2003. , which is further based on the work by the Tavistock Institute[6] T. Institute. "The Tavistock Institute," http://www.tavinstitute.org/.

Related work includes Crowston [11], who suggests identifying coordination problems faced by an organization in order to arrive at alternative coordination mechanisms to manage them. Crowston [11] builds upon the Coordination Theory framework described by Malone and Crowston [12] and present a typology of dependencies and their associated coordination mechanisms. Crowston [11] defines three basic types of dependencies: (i) between task and resource, (ii) between two resources and (iii) between two tasks. He then describes a method to determine coordination problems related to particular dependencies. However, Crowston [11] does not address why a particular coordination mechanism has to be applied for a particular coordination problem, and also, how one can identify coordination problems. While the emerging field of "Socio-Technical Congruence" [13, 14] provides some answers, they focus solely on a particular coordination problem - the Conway's Law STSC (based on resource-resource dependency[11]) and its impact on software development [15]. Through STSCs, we extend this notion of *lack of Socio-Technical Congruence*, by including coordination problems that are a result of dependencies between tasks and those between tasks and resources [11] (as we shall explain later). In this paper we report on a method and a tool that a project manager can use in order to detect particular Socio-Technical coordination problems or Socio-Technical Structure Clashes (STSCs). We thereby extend the work of Crowston [11] by suggesting particular coordination problems (STSCs) and their associated coordination mechanisms (part of the Socio-Technical Pattern), along with a method of identifying the STSCs. We demonstrate a method for monitoring the software development process by using social network data as well as data from the software code and architecture. Such a method would also aid in improving the design of projects (teams and task assignments). We describe the validation of our method in a detailed case study of a software development company called eMaxx. In the case study we demonstrate data collection from multiple sources, including; mining of bug tracker data, mining the version control system and analysing data collected through interviews. The core contribution of this paper is in the identification of STSCs with our TESNA method and tool, which is an extension of current analysis methods [3, 7, 8], through the usage of multiple Socio-Technical Patterns. In this paper the research follows the design science research methodology, as described by Hevner et al. [16]. We hence motivate the problem, define the objectives of a solution, describe the TESNA method and tool and later demonstrate and empirically validate them through an industrial case study (following the guidelines in [17]). In the process of validating the TESNA method and tool, we also provide empirical validation of the Socio-Technical patterns that we use.

The rest of the paper is structured as follows; section 2 covers the theoretical background of patterns, section 3 lists the requirements for a method and tool, section 4 explains the features of the TESNA method and tool, section 5 describes the eMaxx case study, section 6 then describes the results of the case study, section 7 discusses the results along with the internal and external validation and finally section 8 concludes the paper.

## 2 THEORETICAL BACKGROUND

### 2.1 Patterns

We first describe the term "Pattern". Christopher Alexander, who conceived the notion of patterns in the field of architecture, describes a pattern as "a recurring solution to a common problem in a given context and system of forces" [18]. In software engineering, patterns are attempts to describe successful solutions to common software problems [19]. Patterns reflect common conceptual structures of these solutions and can be used repeatedly when analysing, designing and producing applications in a particular context. Coplien and Harrison [7] define pattern as "*a recurring structural configuration that solves a problem in a context, contributing to the wholeness of some whole, or system that reflects some aesthetic or cultural value*" ([7], p14). Patterns represent the knowledge and experience that underlie many redesign and re-engineering efforts of developers who have struggled to achieve greater reuse and flexibility of their software. Socio-Technical patterns deal with the problems arising from a mismatch between the social (as in the social network) and technical (as in the technical dependencies) structures in an organization. These Socio-Technical Patterns are based on Socio-Technical principles (as shown in Table 1). Currently there is a lack of a method and tool to identify the problems posed by these patterns. Our objective in this research is to design, develop and validate such a method and tool to identify STSCs.

A classic Socio-Technical interaction pattern is Conway's Law Pattern (described in detail in the next section) and a lot of literature has been devoted to it [7, 11-16]. However, in our study, we found two other patterns useful in identifying STSCs, namely, the Code Ownership Pattern [20] and the Project Coordination Pattern [21].

We are interested in patterns describing the interaction between developers working on the different software modules (Code Ownership Pattern), how the interaction between software modules affect developers (Conway's Law Pattern), and how the interaction between developers affects the development activity (Project Coordination Pattern). While Conway's Law Pattern and Code Ownership Pattern focus on the interaction between the software and the developers, the Project Coordination Pattern deals with who coordinates the Socio-Technical interaction. The STSCs related to the patterns provide are a representative sample that covers the typology of dependencies that require coordination [11]. While Conway's Law STSC is related to the resource-resource dependency, Code-Ownership STSC is related to task-task and the Project Coordination STSC is related to the task-resource dependency[11].The patterns are described in the format from Coplien [8] Table 1 below. In the rest of this section we describe these three Socio-Technical Patterns in more detail.

**2.2 Conway's Law Pattern**

The classic paper by Conway [22] states "*organizations which design systems (...) are constrained to produce designs which are copies of the communication structure of these organizations*" ([22], pg. 31). In other words, if the teams involved in software production have shortcomings in their interpersonal relationships, the resulting technical architecture of the software is likely to be flawed. Parnas [23] also described modularization as a "responsibility assignment" rather than a subprogram, thus implying that software modules must be assigned to different developers as tasks [23]. These two papers (Conway [22] and Parnas [23]) imply that the dependencies between the software modules should reflect the social communication structure of the Software Team in the ideal situation. This concept of "homomorphism" between the social communication structure and the technical architectural dependencies has come to be known as Conway's Law. Conway (1968) also suggests that as a new architectural design is almost never the best possible one, it needs to be changed and improved [22]. This changing of the architecture implies that the organizational communication must be flexible enough to adapt to this change. Thereby suggesting that the converse of Conway's Law, i.e. that the architecture of the product should drive the organizational communication structure, directly follows (or needs to follow) from Conway's Law [8, 22]. When this fails to happen, the dependencies between the modules would not be supported by communication between the developers. In this research we call such a situation a Conway's Law STSC as described in Table 1. Crowston (1997)[11] in his typology of Coordination Problems, describes the Conway's Law STSC with these words: "…*there are dependencies between modules owned by different engineers, that is, **resource-resource dependencies, that constrain what changes can be made. A module depends on another if the first makes use of services provided by the second*." (Page 167). Hence Conway's Law STSC is related to a resource-resource dependency[11].

Herbsleb and Grinter [24, 25], in their case study of a software development company, show that unpredicted events that are difficult to manage can occur due to coordination problems in a particular project. They refer to coordination problems due to the presence of Conway's Law STSC, and in their words "qualitative data from the case study strongly supports Conway's and Parnas' positions that the essence of good design is facilitating coordination among developers" ([24], p69). Morelli et al. [26] describe a method to predict and measure coordination-type of communication within a product development organization. They compare predicted and actual communications in order to learn to what extent an organization's communication patterns can be anticipated. Sosa et al.[27] find a "strong tendency for design interactions and team interactions to be aligned," and show instances of misalignment are more likely to occur across organizational and system boundaries. Sullivan et al. [28] use Dependency Structure Matrices (DSMs) to formally model (and value) the concept of information hiding, the principle proposed by Parnas to divide designs into modules [23]. de Souza et al. [29] describe the role played by APIs (Application Program Interfaces) which limit collaboration between software developers at the re-composition stage [30]. Cataldo et al.[31] as well as Wagstrom and Herbsleb [32] perform the same study of predicted versus actual coordination in a study of a software development project in a large company project. Their work provides insights about the patterns of communication and coordination among individuals working on tasks with dynamic sets of interdependencies. They define a metric (called *Congruence* – short for Socio-Technical Congruence*)* to measure the proportion of the coordination requirements that we satisfied by a coordi-nation activity, and then see how this metric varies over time. This is also similar to the work done by Amrit et al.[33]. Sosa [34] builds on the DSM based method of Cataldo et al. [31] and provides a structured approach to identify the employees who need to interact and the software product interfaces they need to discuss about. Gokpinar [35] analyse the misalignment of product architecture and organizational interaction and show that greater misalignment leads to a larger number of filed incident report, thus implying lower quality. Kwan et al. [15] extend the work of Cataldo et al. [31] and analyse the effect of socio-technical congruence on the probability of successful software builds. They introduce a new measure of the congruence metric – a weighted one, but find that there not much relation between the congruence measure and the success of a build, for a majority of the builds [15]. They explain that due to the distributed nature of the project, the developers used different means of coordination other than direct communication[15, 36]. Hence, it is quite clear from literature that the effect of Socio-Technical Congruence on software development is dependent on the development context[15, 37]. In a distributed setting what could be vital is to analyse the dependencies among the teams working on the different parts of the software architecture [38]. This can also be the case where the development teams are collocated (in the same building), but in different rooms/floors[39]. In this paper we show how we analysed dependencies at various levels of the software architecture (at the level of the code – syntactic dependencies, and at the level of the system components) and finally used the dependencies among the architectural system components to identify the Conway's Law STSC. Where, at the level of the architectural system components, we are interested in the effective inter team communication rather than the individual inter-developer communication.

**2.3 Code Ownership Pattern**

As described in Table 1, the Code Ownership STSC is based on the Code Ownership pattern [8].

TABLE 1: THE SOCIO-TECHNICAL PATTERNS USED IN THIS PAPER

| Pattern Name | Conway's Law Pattern | Code Ownership Pattern | Project Coordination Pattern |
|---|---|---|---|
| Principle behind the pattern | Conway's Law: Organizations which design systems are constrained to produce designs which are copies of the communication structure of these organizations. Conway (1968) [22] | Code Ownership: The ownership of each software subsystem ensures that the subsystem is someone's or some team's responsibility Nordberg (2003) [20] | Project Coordination: The person who has the appropriate qualifications and traits to coordinate should do the actual coordination Jha and Iyer (2006) [21] |
| Reference(s) for the pattern | Coplien (1994)[8], Cataldo et al. (2006) [31] | Coplien (1994) [8], Mockus et al. (2002) [36] | Hossain et al. (2006) [40], Mullen et al. (1991) [41] |
| Problem: A problem growing from the Forces. (the problem is not context free) | Organization and Architecture are not aligned | A Developer cannot keep up with a constantly changing base of implementation code. | Lack of appropriate coordinator |
| Context: The current structure of the system giving the context of the problem (gives an indication of the current structure of the system and could hint on other possible patterns that can be applied) | An architect and a development team are in place. | A system with mechanisms to document and enforce the software architecture, and developers to write the code | Social Network of the team at different stages of software development |
| Forces: Forces that require resolution (describe the different considerations that need to be balanced in the solution and hence can be considered a part of the problem) | Architecture shapes communication paths in the organization. Formal organization shapes architecture. | Most design knowledge lives in the code; navigating unfamiliar code to explore design issues takes time. Not everyone can know everything all the time. | People who are not central to the software development or management (or not appropriate) take a central role in coordination |
| Solution: The solution proposed for the problem (solution represents the preferred way to deal with the problem based on knowledge from best practice solutions gathered from practitioners and researchers) | Make sure organization is compatible with the architecture | Each code module in the system is owned by a single Developer. Except in exceptional and explicit circumstances, code may be modified only by its owner. | Make sure the appropriate people take a more central role in coordination. |
| Resulting Context: Discusses the context resulting from applying the pattern. In particular, trade-offs should be mentioned | The organization and product architecture are aligned. | The architecture and organization will better reflect each other. | Project critical information will be conveyed to all team members. |
| Design Rationale/Related patterns: The design rationale behind the proposed solution. Patterns are often coupled or composed with other patterns, leading to the concept of pattern language. | Changing interfaces between modules could lead to serious bugs | Lack of code ownership is a major contributor to discovery effort in large-scale software development today. | Coordination done by inappropriate personnel could lead to costly delays in completing the task |

The related problem is that a developer would find it difficult to cope with a changing base of code. The same STSC applies to a situation where a developer who was not involved in the development of a particular code module is suddenly given the responsibility for a particular release of the code. Such a situation requires significant coordination, in order to get the new developer informed of the history of changes behind the current version of the code. This situation can create a substantial coordination requirement, depending on the number of developers who have made changes to the module and on the kinds of changes that have been made until then.
Crowston (1997)[11] describe the shared resource dependency of having to make multiple bug fixes to the same code as an example of a task-task dependency. In his words: "*In this process, this dependency is managed by* *assigning modules of code to individual programmers and then assigning all problems in these modules to that programmer. This arrangement is often called "code ownership""* (page 167). Hence, the Code Ownership STSC is related to a task-task dependency. Nagappan et al. [42] describe a measure of what they call Organizational Code Ownership (OCO) that measures how diverse the contributions to the code are. When the contributors are very diverse, they claim that there could be higher chances of defective code, due to synchronization issues, mismatches, build breaks etc. [42]. The Code Ownership STSC is especially problematic when the developers do not follow an XP methodology with collective ownership guidelines like continuous integration, 100 percentage unit testing and strong coding style guidelines [20]. Code ownership has been widely cited as a coordination mechanism [36, 43]. However, relatively little has been published on the lack of code ownership or the coordination requirements caused by faulty code ownership practices [20]. LaToza et al. [44] in a survey conducted on developers of a software organization describe how developers maintain mental models of the code. They conclude that personal code ownership is usually tacit and written records are considered out of date and ignored. On the other hand they describe a stronger notion of team code ownership among developers [44]. In this paper we describe how one can identify a Code Ownership STSC and show the results of doing so in our case study.

## 2.4 Project Coordination Pattern

The role of a project coordinator has long been acknowledge as being important in software development [45]. However, there is not much literature on the required skills and qualifications for a project coordinator [21]. Kerzner (2009)[46] describes planning, coordinating, analysing and understanding the organization as skill requirements to being a project coordinator. Jha and Iyer (2006)[21] discuss the traits of an effective project coordinator. They find statistically significant differences in the traits of successful versus not so successful project coordinators [21]. They list the key traits for an effective coordinator as human relationship skills, timeliness, technical knowledge of the subject, belief in team playing spirit and coordination ability [21].

The Project Coordination principle suggests that the person actually coordinating the project should be someone who is appropriate to coordinate. Traditional project wisdom suggests that this person should have an overview of the project, have experience or the ability to coordinate, and have the authority to coordinate. It is likely that most communication in the project will go through the coordinator [40]. That is, the coordinator is likely to be central to the project-related communication. However, if the person central to the communication does not have the appropriate qualifications for coordinating the project, for example, he/she does not have an overview of the project, does not have prior coordination experience, or is not in a position to exercise coordination authority, coordination problems can arise. As the judgement of the traits of the coordinating person is a subjective exercise, the Project Coordination pattern leaves this task with the manager of the project. The Project Coordination Pattern only addresses the mismatch between who is supposed to coordinate and the actual coordinating person. Crowston [11] describes task assignment dependency as an example of task-resource dependency: "*Task Assignment is a coordination mechanism for managing the dependency between a task and an actor by finding the **appropriate** actor to perform the task*". Hence the Project Coordination STSC is related to the task-actor or task-resource dependency. Where the *task* is one of coordinating or allocating the bugs and the *actor* is the person performing the task.

In order to determine who has the central coordinating role in the project communication, we first need to determine the appropriate centrality metric. For this we consider popular centrality measures like the Betweenness Centrality and Information Centrality [47]. Betweenness refers to the frequency with which a node falls between pairs of other nodes in the network. In other words, betweenness centrality is a measure of, "*the degree that each stands between others, passes messages and thereby gains a sense of importance in contributing to a solution, .. , the greater the betweenness, the greater his or her sense of participation and potency*" [48].

Freeman et al. [49] perform an experiment where five people (with no previous history of interaction) are placed in different structural positions while enforcing a strict pattern of communication. They try to determine which network positions are most conducive to coordination. From the post experimental interviews, "betweenness" emerges as being the most useful for coordination [49]. This result is further supported by Mullen et al. [41] who state "*the individual in the most centralised position in a network in terms of Betweenness is likely to emerge as the leader…*". They further go on to state "*this indicates that the potential for the control of communication is a critical contribution to the participation in, and satisfaction with performance in communication networks*." ([41], p13).

However, when the information flows through pathways that are not always geodesics (as in the case study that we discuss later), it is more appropriate to use the lesser known Information Centrality [50] metric. According to Wasserman and Faust [47]: "*So it may make sense to generalize the notion of betweenness centrality so all paths between actors, with weights depending on their lengths, are considered when calculating betweenness counts. The index of centrality developed by Stephenson and Zellen (1989) does exactly this.*"([47], pg. 193). The information centrality of the node *i* is then defined by using the harmonic average:

$$IC(i) = \left[\frac{1}{n}\sum_j \frac{1}{I_{ij}}\right]^{-1}$$

Equation 1: Information centrality of a node [50]

Where the information measure $I_{ij}$ between two nodes is defined as the reciprocal of the topological distance $d_{ij}$ between the corresponding nodes, $I_{ij} = 1/d_{ij}$ (when $d_{ij}$ is zero then $I_{ij}$ is considered as infinite for computational purposes, which makes $1/I_{ij} = 0$)[50].

In the case study that we describe in section 5, we consider communication pathways in a bug tracker discussion where the paths of information transfer are not necessarily geodesics. This is the reason why we use information centrality (instead of betweenness centrality) to analyse potential Project Coordination STSC in this research. Furthermore, in this pattern the social dependencies could be as a result of technical workflow dependencies (as in our case study, the workflow/task allocation in the bug tracker) or technical task dependencies. This is unlike the other two patterns in which the social dependencies are strictly caused by direct technical task dependencies. As described in Table 1, a Project Coordination STSC occurs when people who are not central to the software development or management (or not appropriate), over time, take a more central role in coordination. Analysing the change in the information centrality index can give us an idea of how the role of Project Coordinator in the network changes depending on the tasks at hand.

Table 1 describes the three patterns described above in the pattern format used by [7].

## 3. REQUIREMENTS FOR THE TESNA METHOD AND TOOL

Although Socio-Technical patterns have existed for years, their application to real life projects has proven to be difficult. The reason being, that it is very labour intensive for the Project Manager to find the various problems (refer to Table 1) in a particular project, to which the patterns can be applied. The large amount of data related to a company's social and technical networks makes it very hard to identify STSCs in them. Even gathering the data requires substantial effort, as different projects use a variety of tools for communication and coordination.
The objective of this paper is to design and validate a method and tool that can enable a project manager to effectively identify different types of STSCs by analysing the social and technical networks.

Hence, the overall requirements for the TESNA tool are:
- The tool should be able to access and mine the repositories/log files associated with the different tools used by the developers for communication and coordination.
- The tool should be able to find out and/or utilize the technical dependencies at the level of the source code or system architecture.
- The tool should represent the data in such a format, that it is easy for the Project Manager to identify the various STSCs.

Existing tools mostly deal with different aspects of Conway's Law Pattern and can be categorized as follows:
1. Tools that create awareness like FASTDash[51] and Augur [52]
2. Tools that identify code level dependencies like CollabVS[53], Tukan [54] and Palantir [55]
3. Project Exploration tools like Expertise Browser[56], Team Tracks[57] and Ariadne[58]
4. Socio-Technical Structure exploration tools like Tesseract[59], OSS Browser [60] and CodeSaw[61]

As we shall show in the case study of eMaxx, none of the above tools can sufficiently help in identifying the three STSCs described, including the Conway's Law STSC. In response to this, we have developed a method and tool called TESNA (as reported in [3]) that aids us in identifying various STSCs. The method and tool have undergone considerable development since research reported earlier [3]. Specifically, we have improved the technical metrics and visualizations of TESNA to identify STSCs like the Conway's Law STSC (based on architec-tural, syntactic and logical code dependencies) and the Code Ownership STSC, as we shall report later in the case study. However, the primary contribution of this paper is not in presenting the tool TESNA as a new tool to analyse Socio-Technical STSCs, but rather, the aim of this paper is to present the method of identifying the three STSCs described. This method is illustrated in the case study in this paper. We feel that with sufficient modifications most of the tools mentioned earlier will be capable of detecting these STSCs.

## 4 THE TESNA METHOD AND TOOL

Figure 1 represents an overview of the method. The overall Method consists of several steps. First, we assume that the Project Manager has a fairly good idea about the different Socio-Technical Patterns and about which specific pattern or groups of Patterns have to be applied in the current project situation. Next, the TESNA tool accepts input for the Social Network as well as the Software Architecture (and the software code), and the tool provides a visual description of the networks and metrics, based on the Socio-Technical Patterns selected. If the Project Manager identifies an STSC, the manager can decide whether the planned software process model is appropriate or needs to be changed. Alternately, the Project Manager can decide to modify both the social communication network and the software architecture. The project manager can continue with iterations of this STSC assessment/ structure revision cycles until the requirements of the pattern are under control.

Figure 1 indicates two labelled loops, the SN loop (the Social Network loop) and the ST loop (the Socio-Technical loop). The SN loop corresponds to the Social Network Module of TESNA. The Social Network Module reads input about the Social Network, by mining Chat/Mail/Bug-Tracker Repositories.

The TESNA tool then creates visualisations of the social networks and calculates metrics to show the changes of the networks over time. The ST loop corresponds to the Socio-Technical Module of TESNA. The Socio-Technical module reads input on the socio-technical aspects of the software development process.

In short the TESNA method can be described by the following steps:

1. Exploring potential Coordination problems
    1) Selecting the Socio-Technical Patterns that are applicable
    2) Analysing the social network and/or the software architecture using the TESNA tool
    3) Analysing the visualisations and metrics produced by TESNA for the particular Socio-Technical Pattern
    4) Identifying potential STSCs
    5) Optionally, exploratory interviews can also be conducted.
2. Verifying the STSC: Verification is through interviews or with the help of a workshop presentation
3. Identifying the potential reasons contributing to the structure clash: : Interview data and/or a feedback from a workshop presentation are used to get the reasons behind the STSC

Please note that the last part (3rd point above) of identifying potential reasons for the STSC is an optional part of the method. When the person using the TESNA method is the Project Manager the procedure may be slightly different, like replacing the interviews with direct observation etc.

In order to read the technical network, the tool and analyses the source code (currently TESNA can handle java source code).  To identify the socio-technical links, the tool mines Software Configuration Management Systems like CVS (Concurrent Versioning System) and SVN (SubVersion), as well as other repositories like Mantis Bug tracker[62]. TESNA uses various displays to help identify the existence of STSCs related to the social network and/or the software call graph. Both qualitative and quantitative data are used to identify STSCs. Qualitative data is represented as various graphical visualizations of the social and the technical networks. For example; in this research we use the display of the clustered software modules along with the developers as well as the variation of cumulative information centrality. Quantitative data consists of various metrics that the tool calculates to augment the qualitative data (like the Core Periphery Distance Metric described later). On viewing the displays, the manager can decide if any particular identified STSC is problematic and needs to be addressed. The metrics related to the STSC aid the manager in understanding the extent of the coordination problem. The TESNA method and tool can also be used by developers as an aid to the on-going development work. For the method and tool to be effective, the developers should actively mine the different repositories and have a working understanding of the different Socio-Technical patterns. The details of the method, including how the interviews are conducted and analysed is described in the following sections.

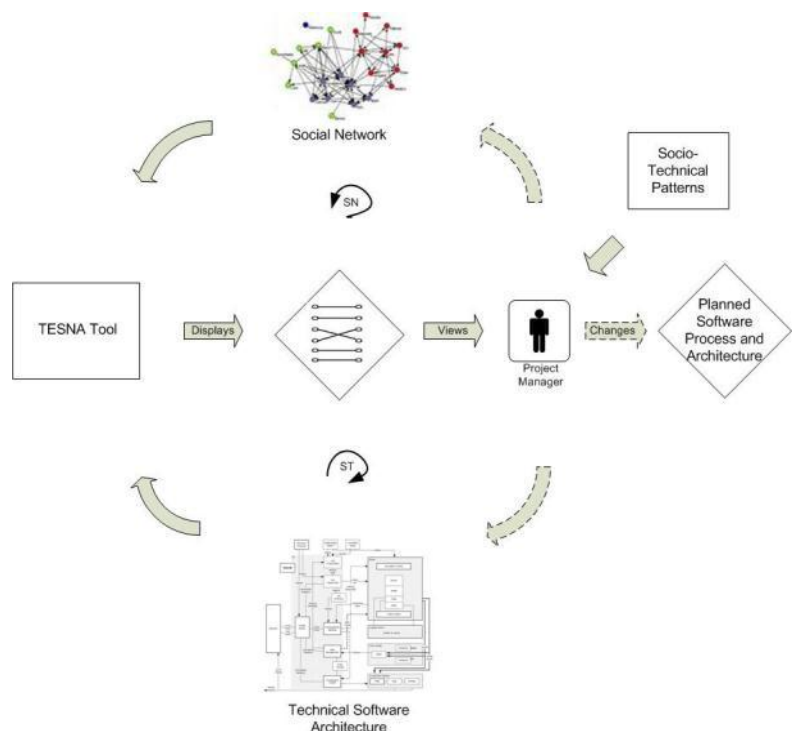

*Figure 1: The TESNA Method and the Planned Software Process*

## 5 CASE STUDY

We used a case study as a means of illustrating as well as validating the TESNA method and tool. We conducted our case study in a software development company called eMaxx. eMaxx is a leading provider of web portals and Mid Office solutions for municipalities in The Netherlands. eMaxx is considered a mid-sized software development company in The Netherlands, and enjoys a 30 - 40 percentage of the market share. The solutions offered by eMaxx are customized for each municipality, so their software development is manpower intensive. eMaxx has around 50 employees including 20 developers. The developers are distributed among three teams. eMaxx follows a variant of the waterfall development methodology. Recently the company has merged with another company called XL21 (this will be visible in the data analysis later on).

The rationale behind the selection of eMaxx for our in depth case study was that (i) a majority of the software development companies in Europe are Small and Medium sized Enterprises (SMEs) [63], and we feel that SMEs have been under-researched (ii) having found STSCs in a small enterprise [3], we intended to see if we could, using TESNA, find instances of the STSCs related to the three patterns (Table 1) in a midsized company, (iii) if it is indeed is possible to locate STSCs in eMaxx using TESNA, then we plan to conduct a case study in a larger corporate environment where there is a greater likelihood of finding both similar and other STSCs related to Socio-Technical Patterns.

Over a period of 6 months we mined the Software Repository (CVS) as well as the Bug Tracker repository (Mantis [62]) used by the developers at eMaxx. We also interviewed the support staff, developers, and the project leaders of each primary software development team, where each team is assigned to a different part of the architecture in Figure 2. For example, the developers working on the Front Office part of the architecture were considered to be a part of the Front Office team and so on. In total we interviewed ten personnel (7 developers, 2 managers and one support staff personnel) covering all the different departments of eMaxx. The demographic data of the interviewees can be seen in Appendix C (Table 2).

Each interview lasted between 1 to 2 hours. We interviewed the employees two times, once before the analysis of the data and once after the data was analysed. The interview before the analysis of the data was done as a means of case study exploration. Among the many questions asked, in this phase, were questions related to what the communication network (as well as frequency) of the employee was, what the modes of communication were and whether the employee had observed STSCs in particular projects. Once the data was analysed we again interviewed the same personnel to understand if indeed there was an STSC. We then conducted a few follow up interviews (in some cases) to find the cause of the STSC. Finally, as a means of validating some of the STSCs, we also presented the analysed data to most of the eMaxx employees (including the ones interviewed) as part of a workshop. In a separate questionnaire given to the participants of the workshop, we gathered feedback for the TESNA method and tool. For the interview process and the subsequent analysis of the data gathered, we used the coding technique described by Miles and Huberman [64].

The architecture of the main Mid Office application and the various teams as well as the task responsibilities of the developers are described in Figure 2. At the request of the CTO of eMaxx, all the names used in this case study are pseudonyms. We analysed the eMaxx architecture (Figure 2) to locate the different dependencies among the architectural components. This was done through interviews of the chief architect – the CTO, with confirmation from the other developers. As seen in Figure 2, the architecture of the primary Mid Office product that eMaxx develops consists primarily of the Front office, Application Server and the BPEL Engine. The business logic in the Mid Office application is modelled using BPEL (Business Process Engineering Language). The business logic is embedded in the Application Server as well as the Front Office, which make them both dependent on the BPEL specification. The Front Office and the Application Server part communicate through XML (SOAP) messages. Hence, the main architectural level dependencies are among the Front Office, Application Server and the BPEL specification.

All the corresponding task responsibilities that changed during the period of the study were duly noted. Though the developers were located in the same building, they were in different rooms and floors. The application server and the front office team were in the ground floor in different rooms (of about 3-5 people each), while the BPEL team was located in the first floor.

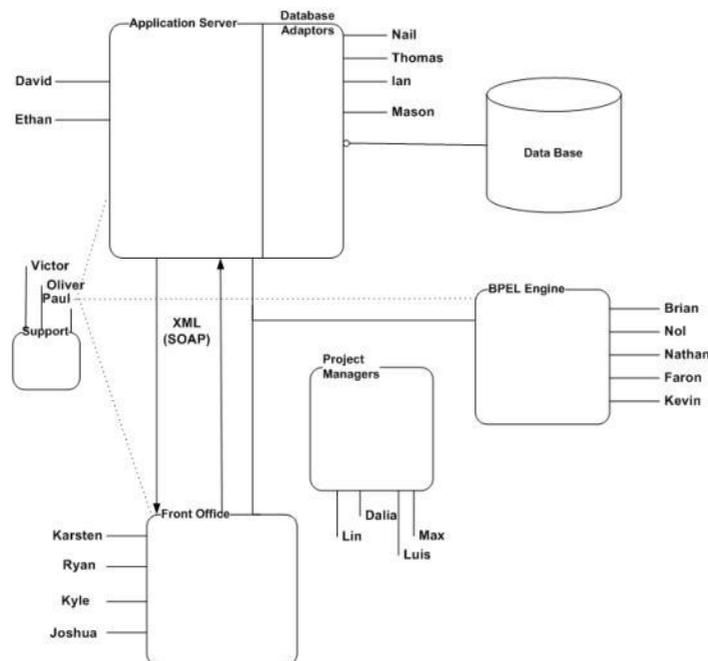

*Figure 2: The Mid Office application Architecture and the task responsibilities*

### 5.1 Mining Different Repositories at eMaxx

We examined three data types from two repositories at eMaxx, namely:
 1) Social Network from the Mantis bug tracker
 2) CVS log data as to who last modified which file in the software application module
 3) Call graph data of the software application modules, also obtained from the CVS

Here, we will describe each data type in more detail. eMaxx uses the Mantis bug tracker [62] to keep track of the progress on software bugs specific to different projects. When asked, the CTO as well as the other interviewees informed us that the bug tracker discussion was an accurate representation of all the project related communication. This included the bug finding and fixing activities but also the coordination activities required surrounding it.

We mined 2250 pages from the Bug tracker corresponding to 23 different customer oriented projects. The bug tracker pages spanned a period of four years, starting from 2004 and ending in the beginning of 2008. Each web page of the bug tracker dealt with a specific bug concerning a software component of a project. From each page, we extracted the sequence in which the developers tackled the particular software bug in question. Data on the names of the developers, the dates at which they sent the messages as well as the names of the projects, were mined with the help of TESNA. The social network of the people who posted messages on the bug tracker was built as follows: if developer A posted a message and developer B replied to it, then a link was established from A to B, similar to the method used by Howison et al. [65]. In this way, we arrived at one social network from each bug tracker page. We then calculated the information centrality of the networks and plotted the change of the cumulative information centrality of each individual involved in a project, over time.

eMaxx uses CVS (Concurrent Versions System) to keep track of eMaxx' s Mid Office application code and Front Office code. However, they used different servers for each (while the BPEL specification is maintained only through flat files). We used TESNA to mine the Mid Office CVS (as it contained the code modules in which the core logic and operation of their software resided) in order to retrieve data on the name of the software class file which had been modified, the name of the developer who modified the file as well as the date at which the file was modified. We wrote a script to check out (from CVS), compile and then used TESNA to analyse all the versions of the compiled source code (jar files) for each of the core Mid Office modules.

We used TESNA to analyse and display the data and functional related syntactic dependencies (call graphs) of each of the application modules (jar files). However, we could not analyse the logical coupling, using the method suggested by Gall et al.[66] for the following reasons: (i) the number of versions in which the individual software classes of the modules were changed, were very few, about 2 on an average. Hence, we could not guarantee the logical coupling between their individual modules [66], and (ii) We could not establish a link between the CVS change reports and the bugs listed in the Mantis bug tracker. As the change reports on CVS were not very detailed, the calculated logical coupling would not be reliable [66]. We could then conduct a temporal analysis of the call graphs, to determine changes in the responsibilities for different versions of the application modules. In total we analysed 29 application core modules, each belonging to the Application Server part of the Mid Office architecture (Figure 2). Of these 29 application modules we selected 14 based on the following criteria: (i) that the modules were worked on by at least two developers, (ii) the modules had at the very least 100 java classes and, (iii) we also selected those modules that were considered important and core to the functioning of the Application Server. We analysed the core-periphery nature of these 14 application modules (as we shall describe later). We determined which developer modified (and as a result was responsible for) which part of the application module and when.

**5.2 STSCs in eMaxx and Feedback**

As explained earlier, we identified three primary classes of STSCs based on the three Socio-Technical Patterns described in Table 1. We used the communication link data from the bug tracker along with the interview data and compared it with the dependencies between the software modules at the level of the code and between components at the level of the architecture, to identify the Conway's law STSC. On the other hand, the Code Ownership STSC was identified by mining the software repository (CVS) used in eMaxx and by determining the ownership of the modifications done on the software over time. We mined the Mantis bug tracker in order to determine the project-specific communication links. We compared and verified these social networks to the ones we got through the individual interviews. The data from the Bug tracker after verification with the interview data was used to identify the Project Coordination STSC. In each case, we not only identified the STSCs, but we also identified the STSCs over time using historical longitudinal data.

After the data was analysed, we took the data back to the CTO (David) for his feedback. We used this feedback as one of the means verifying the STSCs. We also verified the STSCs by asking the project leader from the Support Team, Oliver for his feedback on the data. We followed up on the data that we got from observing the Project Coordination STSC. By interviewing the different people involved in the particular STSC (after the data was analysed and the STSC was identified), we tried to arrive at a better understanding of the reason behind the STSC.

Finally as described earlier, in order to validate the TESNA method and tool, we deployed a questionnaire after presenting the data to the workshop consisting of developers, manager and support staff of eMaxx.

We now briefly explain the algorithms and methods we employed to identify the different STSCs at eMaxx.

**Conway's Law STSC**

The Conway's Law STSC (Table 1) we identified at eMaxx involves technical dependencies at the level of the system architecture (Figure 2) that are not satisfied by corresponding communication between the people involved with the technical artefacts.

We analysed the technical dependencies at the level of the software application modules (with an analysis similar to the one done by Cataldo et al. [31]) as well as, at the level of the system architecture. Through the interviews as well as onsite data gathering, we found that the dependencies at the level of the software application modules were met by face to face communication (team members located in the same room) or communication via chat and e-mail (in the case of team members located in different rooms). Furthermore, just as Mockus et.al. observed about the Apache project in [36], the eMaxx developers used alternate coordination mechanisms and kept the core modules small. Smaller core modules require fewer developers to develop and maintain them, reducing the need for extensive coordination among the core module developers. The developers or testers who worked on the same software application module, were located nearby (shared the same room) or had access to a quick and reliable communication route thereby reducing the need for coordination through communication. Hence, we concentrated our analysis at the architectural level of technical dependencies and the communication among teams, working on the different parts of the architecture (Figure 2). The technical dependencies at the level of the architecture were found by gathering information on the architecture and through the exploratory individual interviews. We found two architectural dependencies that were critical to the process: the dependency between BPEL and the Front Office, and that between BPEL and the Application Server (Figure 2).

**Code Ownership STSC**

As described in Table 1, the Code Ownership STSC is related to the problem that a developer finds it difficult to cope with a changing base of code[8]. In the case study at eMaxx, this coordination requirement is greater than that arising from different developers working concurrently on the same code. This is due to the fact that the dependencies among people working on the project at any given time were met with the required communication (as described in the previous Conway's Law STSC).

In order to detect this STSC we mined the CVS server of eMaxx. After mining the log history and source files, we calculated the call graphs of the software packages from the source files. We then clustered the call graphs using the dependency based clustering algorithm (Appendix A) used by McCormack et al.'s (2006)[67]. Clustering the call graph makes the detection of Code Ownership STSC easier, as otherwise the call graph can be very large, and as a result the bipartite graph of the developers modifying the source code files can get very complex [67, 68]. Also, modularity of a system can be viewed as a cluster of source-code files, as the structure of dependencies in an unit determines its quality [69]. We used the DSM (Dependency Structure Matrix) clustering algorithm[67] over other clustering algorithms as it clusters based on the number of dependencies between software modules (described in the Appendix A). So, when a developer modifies a file in a particular cluster, depending on the kind of modification, all the dependent files that need to be altered as a result of the modification [31] would lie in the same cluster. The clustering together with the calculation of the Core Periphery Distance Metric (CPDM) (Appendix B) simplifies the detection of the Code Ownership STSC as we shall show in the case study that follows.

Once the manager finds a Code Ownership STSC at the level of the clusters the manager can then zoom into the clusters to see the Code Ownership at the level of the source code files and decide if the problem is indeed severe enough to justify action.

**Project Coordination STSC**

As we explained in the previous section, the social network diagrams were grouped based on the projects where the bugs occurred. We could then calculate the information centrality of the people involved in the bug tracker of each project, over time. In the case of the usage of the bug tracker at eMaxx, we analysed the messages posted and observed that most of the coordination work was the allocation of bugs and in routing the replies to the attention of other developers. This had to be done, as the developers at eMaxx did not respond to a bug report unless and until the message was addressed to them. The messages were mostly task assignments (resembling package transfer), that were routed mostly through shortest paths or geodesics ([70], pg. 63 table 2). The reason they were routed mostly through geodesics was that as the personnel knew each other they tried to transfer the responsibility of the bug fix to the right person in the smallest number of steps. However, as explained earlier, since we cannot be sure that the messages were routed through geodesics, we use the information centrality metric rather than the more popular betweenness centrality metric [47].

We could then see who had a higher information centrality at various periods of the project. Depending on whether this is different from what was planned in the project planning stage or expected with the job description, we would have an STSC (Table 1). The 2250 social network diagrams (as explained in the previous section) represented 23 different projects. Among these projects, we found three particularly interesting as they involved more developers than the others and were spread across a comparatively longer timeline. The interesting projects were LPOC, LR and TRM (all pseudonyms). The employees included XL21 employees as well as employees from the customer side (the municipalities involved in the project). In order to increase the readability of the cumulative information centrality images, we decided to include only those persons who had a large enough average information centrality (we ignored the people who had an information centrality close to zero or an insignificant cumulative information centrality).

## 6. RESULTS

### 6.1 Conway's Law STSC

We also identified the social network of the developers, support staff as well as project leaders through exploratory open ended interviews. We asked each developer whom they spoke with (i.e. other developers, support personnel and project leaders) and how often. We also asked if they had encountered any coordination problems/inconsistencies in the projects they were working on.

One of the key developers: Ryan of the Front Office Team had this to say:

*"There is a communication problem between teams..., between Front Off and the BPEL team, they [BPEL team] decide things they should not decide.., they should ask us"...*

*"I think the communication within teams is more than between teams"...*

*"Sometimes they [BPEL team] decide things that are difficult to do with the framework, but most of the time BPEL [BPEL Team] just decides process specific logic, with process specific logic it's easier to accept things". "Sometimes they decide things that require changes in the framework, which is more difficult"…"When they [BPEL team] communicate with us [Front Office Team] it's about standard, not process specific things"*

These statements from Ryan suggest communication related coordination problems between BPEL and the Front Office and Application Server) teams.

The framework that the developer mentioned above refers to the core data structure implemented in the Front Office application based on the BPEL process specifications. So every time the BPEL process-specific logic dealing with the framework changed, the particular Front Office developers (like Ryan) had to be contacted. The technical dependency between Front Office team and the BPEL team made the lack of a smooth communication link between the Front Office team and the BPEL team a coordination bottleneck. This point was also brought up by another Front Office developer, Kyle, who had this to say:

*"There were problems between BPEL [team] and Front End [team], because BPEL did things differently than before...*

*So you develop things and you think that 'now I can communicate with BPEL [to the Application Server]', but it doesn't work; all I have are errors and faults...*

*Then you walk to the BPEL team and they say 'Oh we do it differently now, we have changed it'"...*

*The whole communication within this company should be much better. A lot of problems in this company are related to bad communication!"*

Thus for Kyle, when the BPEL team makes changes to the business logic used by the Front Office applications, they do not inform the Front Office development team of the changes. This lack of relevant communication and transfer of important information, in spite of the existence of a dependency between the teams causes problems in the development process.

This communication problem was also mentioned by Project Manager Lin:

*"..When the clients find an error, we have the Bug Tracker to report the errors, they always go to BPEL first, mostly, they [BPEL team] look at it, but they look at it by themselves, they don't go to the Mid Office [Application Server team], Front Office [team] to discuss and look at the problem and how they can fix it, they [Mid Office team] do it just by themselves"...*

*"So what I did after a while, when I noticed that is that I went to BPEL [team] and then I went upstairs to the Mid Office [Application Server team] and the Front Office [team] and I put the three together to solve the problem, because they don't do it by themselves"...*

*"Some people do it by themselves, but most people think 'Oh well it's not my problem! It's the Mid Office [team's] problem or the Front Office [team's] problem'. Many times it's the interaction between the two where the real problem is.". "In the beginning, I thought that the two people know it's a problem between them, and they will solve it together, but I realised it's not always the case"*

When asked, which team, in her opinion, was problematic, she noted:

*"It looks like the problem lies with the BPEL guys, because the problem comes to them and they have to look, as they are the middle ones [in the architecture] to see who can solve it. Whether it's a Front Office [problem] or a Mid Office [problem], they have to see who can solve the problem. Most of the time they say 'It's your problem you solve it', but it [the company] is a team so we all have to solve the problem together."*

These statements reiterate the Conway's Law STSC at the architectural level caused by the BPEL team's lack of communication with the other teams.

On realising that most of the dependencies were between teams, we calculated the number of messages between teams for each social network (2250 pages of the bug tracker). We calculated the number of messages over time between BPEL and Front Office as well as between BPEL and the Application Server team and plotted the distribution as a box and whisker plot [71] in Figure 3 (plotted as InterDepComm). To get a better feel of the lack of communication between these teams, we compared this distribution to the distribution of the number of messages between the Support and the three teams (Front Office, Application Server and BPEL), plotted as SuppDepComm in Figure 3. We finally compare both distributions to the distribution of the total number of messages between people participating in the bug tracker, plotted as TotalComm in Figure 3. From Figure 3 it is clear that the distributions InterDepComm and SuppDepComm, have a median of 0 and SuppDepComm has a higher maximum than InterDepComm (represented by the highest circle, or outlier of the box plot), implying a few networks with more messages. While both distributions pale in comparison to the distribution of the total number of messages among the people using the bug tracker. This result shows that there were few task assignments or enquiries between BPEL and Front Office as well as between BPEL and the Application Server team, when this was essential at eMaxx. This result confirms what the Project Manager, Lin told us, namely, that even though the BPEL team got the bug report from the customer, they did not assign or discuss it with the other teams through the Bug Tracker.

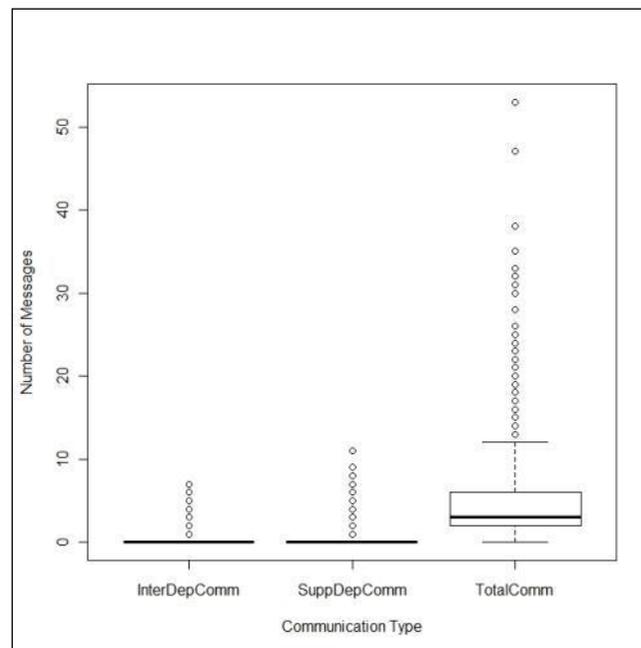

*Figure 3: The comparison of the three distributions of number of messages in the bug tracker over time.*

**FEEDBACK**

To follow up on the analysed data, we decided to try and understand the source of the communication problem with the BPEL team. With this in mind we interviewed one of the main developers in BPEL, Faron. When we asked him who he talks to in his team (BPEL) he said:

*"I spoke to no one for the last six months; now and then I talk to the Team Leader [Brian]"..*

*"Now it's better as I am in the same room"*

These statements clearly show that Faron communicated only with those in close physical proximity. On other occasions he didn't take the initiative to do so.

When asked who he communicated with in the Application Server and the Front Office teams:

*"I worked on a lot of projects, a lot of it on my own and with some mid office people like David"...*

*"I discuss problems and new functionality with David"...*

*"I do the proof of concepts; just make the functionality, not a common way of doing a project,*

*It's faster and has less documentation"*

These statements suggest that Faron mainly spoke with David in the Application Server team and only when he had a problem or a new functionality to discuss. As the project manager Lin told us, when the clients encountered a bug in the project they contacted the BPEL team directly. We also found that Faron was contacted by the clients on multiple occasions. So though the other teams like the Front Office team were waiting for data regarding BPEL changes, or bugs from Faron, he was not aware of it or didn't take heed of this. Identical behaviour from the rest of the BPEL team caused the Conway's Law STSC to occur.

### 6.2 Code Ownership STSC

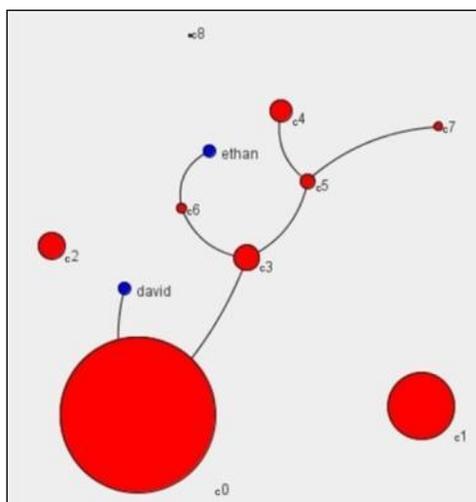

*Figure 4: The software clusters of MZM version 1.7.2.2 along with the developers, David and Ethan who modified the classes in the different clusters.*

In Figure 4 we notice the software clusters of MZM (labelled c0 to c8), an application module in the Application Server along with the developers who modified the different clusters. Here the size of the cluster indicates the number of dependent source files in the cluster. The larger and more connections a cluster has, the more we consider it to be Core. We then calculated the CPDM of the developers (Appendix B) from this clustering process and displayed it as a graph. From this graph we can elicit the Code Ownership STSC. The CPDM ranges from 0 which represents the most periphery part of the application code module to 9 which represents the most core part of the application module. A higher CPDM implies that the developer modified the core part of the software. While a low CPDM indicates that the developer modified the periphery of the software. In Figure 4, we see the 1.7.5 version of the MZM application module, which is one of the main application modules of the eMaxx Mid Office application. In the figure we see David working on classes in cluster c0, while Ethan works on classes in cluster c6. As cluster c0 is more central (in terms of size and connectivity to cluster c6), we find the CPDM of Thomas to be higher than David as shown in Figure 5.

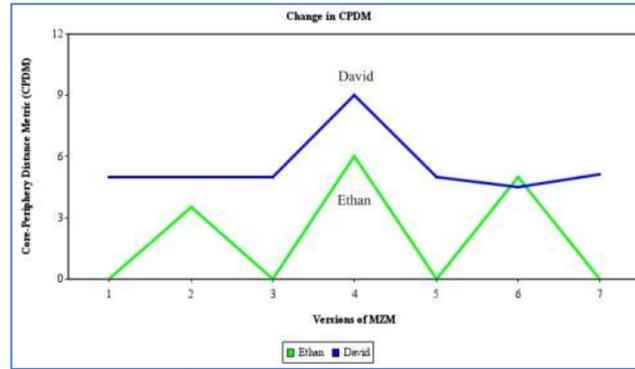

*Figure 5: The Core-Periphery Shift of the MZM application module from the Application Server*

Figure 5 describes the variation of the CPDM of the different versions of MZM. The versions of MZM started with version 1.7.1 and developed to version 1.7.3.2. The versions also represent the timeline of development of the software, i.e. 1.7.1 was developed before 1.7.2 and so on (these also correspond to the versions in the CVS). From the figure we can arrive at the conclusion that there are two developers creating and modifying the application module, namely David and Ethan. In Figure 5 we see that David has a CPDM of 5 while initiating the project and working on version 1.7.1 of MZM, while Ethan has a CPDM of 0. This implies that while David was working on the core part of the software Ethan was not modifying the software at all. While for version 1.7.2 we see that David's CPDM hasn't changed, Ethan's CPDM is 3.5.

This shows that in this version Ethan modified a reasonably core part of the software. What we can see from the trend of this graph is that from version 1.7.1 to version 1.7.3.2 David consistently has the highest CPDM indicating that he is modifying the most core part of the software module. Thus we see that David has Subsystem Code Ownership [20] of this software application module.

Having seen a positive example of work practice, let us now consider a counter example that is also an instance of an occurrence of Code Ownership STSC.

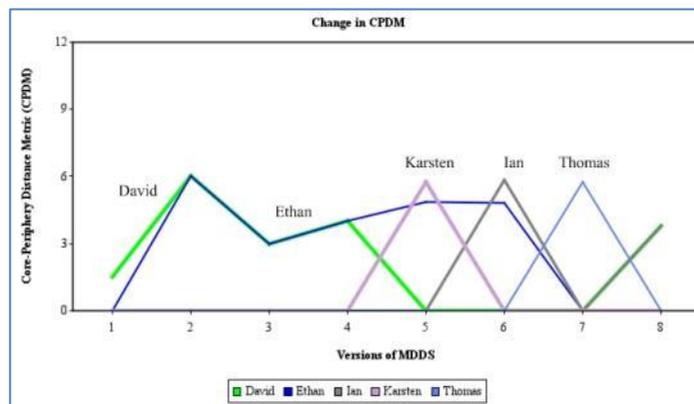

*Figure 6: The Core-Periphery Shift of the MDDS application module from the Application Server*

Figure 6 shows the variation of the CPDM for the different versions of the MDDS project. In this project we see that David and Ethan were involved in working on the core of the software at the start, as they have a relatively high CPDM until version 1.7.5. Then David stops working on the software module as his CPDM drops to zero in version 1.7.6, while Ethan continues to work at the core of the software till version 1.7.7. In the meanwhile we find Karsten and Ian working on the core of the project (having a CPDM of nearly 6), on versions 1.7.6 and 1.7.7 respectively. This doesn't cause a very high coordination requirement (nor does it constitute a Code Ownership STSC) because we also find Ethan working on the core of the software (having a CPDM of nearly 5) during the same period. So it's easy for Ethan to discuss the details of the code with them. While on the other hand in version 1.7.8 we find only Thomas working on the core of the project with a CPDM of nearly 6. This places a very high Coordination Requirement on Thomas, as he has to discuss the details of the changes that David, Ethan, Ian and Karsten have made if he is going to be modifying dependent files as is quite likely given his high CPDM. Hence this large Coordination Requirement makes this work practice a clear candidate of the Code

Ownership STSC.

On presenting the results to the developers, we found that they agreed with the STSCs, especially David, who is also the Team leader of the teams involved in the projects studied. David said that the data on the MDDS module presented does show the presence of a Code Ownership STSC, as Thomas did need to discuss the programming details with the other developers.

In this case study we were able to identify the Code Ownership STSC at the level of the package application module without having to analyse the code ownership at the level of the source code files. Also, the developers knew that the code they had worked on was indeed dependent (according to the data gathered in the feedback session). In other settings, depending on the nature and extent of code modifications, the project manager might want to analyse the Code Ownership STSC at various levels, like at the level of the source code files.

**6.3 Project Coordination STSC**

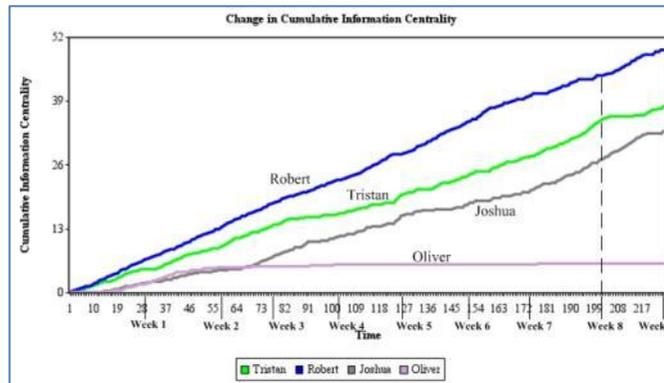

*Figure 7: The variation of cumulative Information centrality for the four most active people in the LPOC project.*

LPOC and LR were two phases of the same project. While LPOC was the initial design phase, LR was the implementation phase. When qualitatively analysing the cumulative information centrality graphs one has to keep in mind that the greater the slope of the graph greater the change in the person's information centrality, and hence greater is the involvement of the person in coordinating the project in the observed duration.

In Figure 7 we see the cumulative Information Centralities of the four active personnel in the LPOC project. In this project there were 24 employees involved, of whom 6 were from eMaxx and the rest from the customer side. Again we asked the CTO, who had the central role of coordinating this project. The CTO mentioned a project leader from eMaxx. As there were no project leaders from eMaxx participating in the Bug Tracker discussion, we would expect the eMaxx support staff (Oliver) to take over the role of coordinating the discussion in the Bug Tracker. When we analyse the information centrality of the LPOC project we see Robert, who is the project manager for the customer, taking over the central coordinating role in the Bug Tracker discussion of the project from its inception to around week 8. Around week 8 we see Joshua, who is in the Front Office team taking over. Again, since, for most of the project there are no employees from eMaxx support team involved in the Bug Tracker message coordination, we can conclude that there is a Project Coordination STSC for this project.

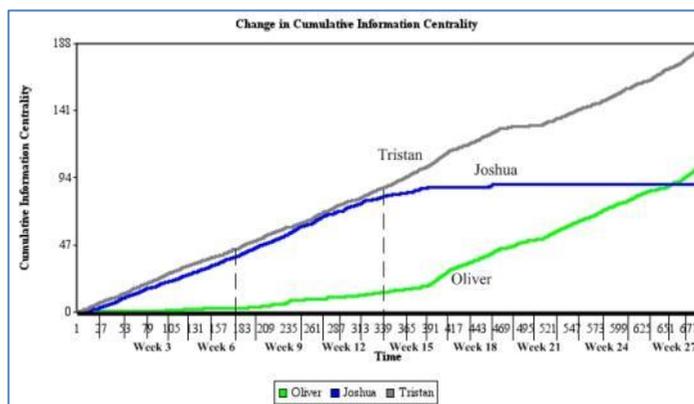

*Figure 8: The variation of cumulative Information centrality for the three most active people in the LR project*

In Figure 8 we see the variation of cumulative Information centrality of the three most active personnel involved in the LR project. The LR project had 34 employees participating in the Bug Tracker discussions, 15 of whom were from eMaxx. Among employees from eMaxx who participated in the Bug Tracker, 2 were Project Leaders (Gavin and Luis), 2 support staff (Oliver and Paul), 4 employees from the Front Office team (Joshua, Karsten, Ryan and sander), 4 employees from the Application Server team (David, Ethan, Thomas and Ian) and 3 employees from the BPEL team (Faron, Brian and Nathan).

When we asked the CTO of eMaxx about who had the main coordinating role in the LR project, the CTO mentioned 2 project leaders (Gavin and Luis) as well as one Front Office member namely Joshua. Oliver, who is part of eMaxx support, said that the main coordinating role for the project in the Bug Tracker was filled by Joshua, Luis and himself (Oliver). On analysing the change in cumulative information centrality of the LR project, we notice that Tristan, (belonging to the customer side of the LR project) has the central coordinating role in the discussions in the Bug Tracker from the inception of the project (week 0) to around week 6. Around week 6 Joshua takes over the coordinating role (as expected) till around week 14. Then from week 14 the central coordinating role in the Bug Tracker discussion is again taken up by Tristan. As Tristan was not supposed to take the central coordinating role, we see a clear case of a Project Coordination STSC.

Figure 9 shows the change in the cumulative information centrality of the three most active people involved in the TRM project. The TRM project involved 9 employees; 4 from the client side (Peter, user30, user 33 and user14), 2 from the application server team (David and Ethan), and 2 from eMaxx support team (Victor and Oliver) and Max, the Project Leader from eMaxx. On calculating the information centrality for the project we see that Max has very high information centrality throughout the project. When the CTO was asked who had the main coordinating role in the TRM project, the CTO mentioned himself and the CEO. This clearly shows a discrepancy from the expected coordination role to the actual coordination.

**FEEDBACK**

We followed up on the LPOC, LR and the TRM projects where we found a Project Coordination STSC. We interviewed the project leaders and developers involved in the coordination of these projects in order to understand why they undertook the coordination responsibility in these projects.

In the case of the LPOC project (Figure 7) we interviewed the Team Leader Robert from the client side (a customer company of eMaxx). We asked him why he took up the coordination of the LPOC project though he was actually a client. He said that the client in this case had a better understanding of the business process involved in the project, so it made sense to coordinate the project. Also, the fact that he knew the technicalities of the project quite well helped him take up the coordinating role. We think that Robert's role in coordinating most of the bugs in the project is still an STSC, as it's not a good idea to rely on the client to manage the software development and testing process. Furthermore, such a situation was unintended and the eMaxx management was not aware of it.

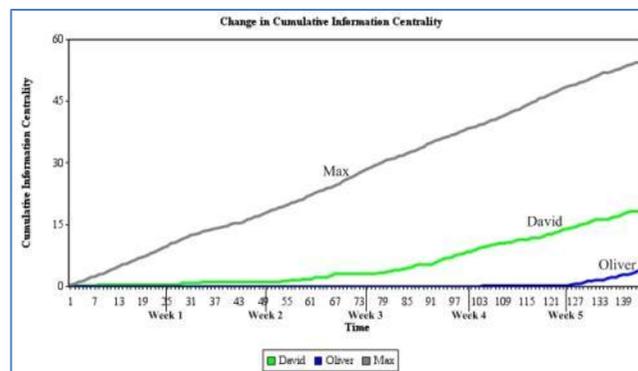

*Figure 9: The variation of the cumulative Information centrality for the three most active people in the TRM project*

In the case of the LR (Figure 8) project, we interviewed Tristan who was the Project Manager from the same customer site as the LPOC project. We asked him why he took the coordinating role and how difficult it was to coordinate the bug reports as well as the project as a whole. The reason Tristan gave was that no one from eMaxx took up the responsibility to coordinate the project themselves. This, combined with the fact that they knew exactly what technical specifications they wanted, made it easier for him to coordinate the project progress and also the interaction in the bug tracker. However, he had reservations on the way the support personnel at eMaxx were structured. He preferred it if there were different people for the roles of project management, support (in the bug tracker) as well as the testing of Bugs. Implying that eMaxx was short staffed and this could be one of the reasons that persuaded the customer side to take up the coordination responsibility. Furthermore, he was clearly not pleased to have taken up the coordinating responsibility. We confirmed the remarks of Tristan when interviewing Max who was one of the Project Leaders at eMaxx.

The project leader of the support team, Oliver confirmed these statements. When asked which team is not so fast to respond to bugs in the Bug Tracker, he had this to say:

*"Our own team is the biggest problem because Paul and Victor [pseudonyms used] have too little time...*

*Also projects can have a problem with that, as Paul and Victor have 70 to 80 hours a week planned. We have to change it back to 40, but of course somewhere we will have a problem with that.*

*I think it is a good idea to split things up before the project is completed and not after delivery."*

This confirmed the fact that the support team was indeed overworked and understaffed, leading to customers taking the initiative to handle the coordination.

In the case of the TRM project (Figure 9) we asked the project leader Max why he had coordinated most of the project in the bug tracker. Max explained that the project management of the TRM project was assigned to an external company CG. As the personnel in the company didn't understand the technicalities of the project someone from eMaxx had to take up the responsibility to coordinate the technical part of the project. As the main project manager (also the CEO of the company) was very busy at that time, Max took up the responsibility himself. In his words:

*" ..They (CG employees) did not have the right understanding of the technical details. They did not have the knowledge to understand the change requests, and hence Project Management becomes difficult. It was very difficult for them to understand or change things."*

Thus implying that, because the Project Managers had not much insight or knowledge of the project he had to take the initiative of managing the project himself.

## 7. DISCUSSION

Earlier we stated that identifying mismatches between software and project team structures is difficult in a dynamic and iterative environment. In dynamic environments, the technical structure of the code changes quite rapidly and so too does the Socio-Technical structure of, i.e. who is working on which part of the code. Similar problems occur in projects requiring rapid iterations, where the person managing the current iteration does not coordinate with the person who managed the previous iteration. We observed instances of both these mismatches in the eMaxx case resulting in the Conway's Law and Code Ownership STSCs respectively. The cause of the Conway's Law STSC could be the lack of communication due to the physical location of the teams as in our case study the BPEL team was in the different floor and hence isolated from the front and application server teams [39]. On the other hand, the cause of the Code Ownership STSC was clearly a lack of explicit code ownership. On the other hand, dynamic and iterative projects can also lead to employees assuming different social responsibilities, if and when they find that their co-workers have not assumed their own responsibilities. Such a situation gives rise to the Project Coordination STSC as seen in the eMaxx case study. In the eMaxx case study we have seen that in the LPOC, LR and TRM projects Robert, Tristan and Max had proactively taken up the coordination responsibility. Both Robert and Tristan were from a customer company of eMaxx and at least one of them (Tristan) was not very pleased to take up the coordination responsibility. A potentially appropriate coordinator could be the main developers who are the key source of technical and organizational knowledge [72].

It is clear from the case study that identifying the three types of STSCs is not easy and is not possible with the support of software tools and methods (described earlier in section 3). The reason behind this is that in the case of the Conway's Law STSC, one has to identify the data dependencies among the teams working on the different parts of the architecture (this is not trivial as the XML data, in this case was not stored in a repository). In the case of the Code Ownership STSC, current generation of tools and methods [36] do not provide a simple longitudinal description of the Code Ownership for different parts of the software binary. While in the case of the Project Coordination STSC, existing tools do not aggregate and display the longitudinal variation of the Information centrality. Earlier, we had described the two requirements of mining and data analysis for the TESNA method and tool. We have described through the eMaxx case study, how the TESNA method and tool could mine the different repositories, analyse and display the data, in such a manner that it was relatively easy to identify the STSCs.

**Contributions**

There are three Socio-Technical Patterns mentioned in the paper. Two of the patterns (Conway's Law and Code Ownership Patterns) are taken from pattern literature, while the Project Coordination Pattern is based on Project Management and Organizational behaviour literature. We have firstly expanded on the definition of Lack of Congruence [73], which is also related to the notion of Congruence gaps [74] and Coordination gaps [35]. We define an STSC as the mismatch of the social network of the software development with the socio-technical dependencies in the software architecture. This includes (i) mismatch of the social network and the social dependencies (due to technical workflow dependencies) - covered by Project Coordination STSC, and (ii) mismatch of the social network and the technical dependencies - covered by Code Ownership and Conway's Law STSC.

Our paper makes a novel contribution in the type of STSCs that the patterns address, and the way the TESNA method and tool supports in the identification of the STSCs. As the paper suggests, in order to apply the pattern, it is not only sufficient to have an understanding of the pattern, but also the different kinds of STSCs one can find related to the pattern. In this regard, the paper makes several novel contributions. Here we consider each of the three patterns and the corresponding contributions they make, both in terms of theory and method:

a) Conway's Law Pattern: Software Engineering literature has described the importance of sharing architectural knowledge and dependencies for the proper coordination of software development [24, 38]. However, most of the previous literature that consider the Conway's Law pattern analyse the syntactic and logical dependencies at the level of the code modules (class files, packages etc.). We think that in a corporate setting, when different teams work on the various parts of the architecture, managing the technical dependencies between the different parts of the software architecture is critical to the development work. In such situations, we hence find an architectural level Conway's Law STSC to be quite critical in determining the success of different software projects [38]. Furthermore, in the eMaxx case study, the STSCs at the level of the software code modules (class files, packages etc.) were more easily fixed (through email/face to face chat) compared to the STSCs at the level of the architecture – which required a more comprehensive discussion among the teams involved. In the paper we discuss an STSC at the level of the software architecture and show evidence for it from our case study. The architectural dependencies are not covered by code level syntactic and logical dependencies, for the following reasons (i) syntax analysers do not cover architectural dependencies like RPC or SOAP calls, and (ii) logical dependency analysis would not include architectural dependencies (like SOAP or RPCs) in the eMaxx case, as the key problem was that, even if the different code repositories were consolidated into one - the Front Office – Application Server and BPEL interfaces were not changed at the same time. Hence these critical architectural dependencies would not show up as local dependencies. So, given the architectural dependencies (these did not vary), we looked into the work related communication between the teams working on the different parts of the architecture over time. We think this method is easy to implement (for a Project Manager) in order to quickly spot such an STSC at the level of the software architecture. Once a Conway's Law STSC is found at the level of the system architecture, one could drill down deeper to identify the STSC at the level of the individual software modules through syntatic and logical dependencies. However, due to the small team sizes at eMaxx (who were also collocated – in the same room), we found that such a Conway's Law STSC at the level of the software modules, involving inter-developer communication did not exist at eMaxx.

b) Code Ownership Pattern: Previous literature addressing code ownership [36] has looked into determining whether a certain level of code ownership existed in projects. However, there is not much literature addressing the Code Ownership STSC, i.e. "A developer cannot keep up with a changing base of implementation code". In our paper we consider this particular STSC and address it with a novel method, thereby, we provide a means of validating this pattern. In order to analyse the STSC related to this pattern, we clustered the dependencies in the software code for a particular module and then determined who worked at the core of the module (through the CPD metric or CPDM) and as a result had a greater ownership of it. We then looked at the variation of CPDM to identify when a developer has to "keep up with a changing base of implementation code", and hence has a higher coordination requirement. We think this is a novel way to identify the Code Ownership STSC.

c) Project Coordination Pattern: Though the importance of an appropriate project coordinator has been discussed in literature [21, 40, 41], there is very little literature on identifying the coordinator of a project and determining if the coordinator is the appropriate person. In our paper we introduce the Project Coordinator pattern to analyse who the coordinator of the project is, and compare it with who was supposed to coordinate the project. In order to analyse this STSC, we calculated different centrality metrics of the work related social network of the people working on a particular project. We then plotted the particular metric (the information centrality metric) in order to identify this STSC.

As in most case study research, the case study presented in this paper can have threats to the validity of its findings. Four threats to validity are considered here: construct validity, internal validity and external validity and reliability [75].

**Construct Validity**
Construct validity addresses the meaningfulness of the results [76]. In order to show that a variable has construct validity we need to show that the measurements are consistent with the intuitive ordering of entities with the attribute of interest [77]. Here we consider the construct validity for each of the type of STSCs found:

Conway's Law: This identification was based on qualitative data and manual identification. There are two measurement variables to be considered for Conway's Law, namely the technical dependencies and the social network of the employees. Ovaska et al. [38] and later Bass et al. [78] have shown that architectural dependencies are important as a means for coordination. Bass et al. also describe the Conway's Law STSC in a broader perspective as architectural misalignment [78]. With regard to the social network, we initially asked the CTO which communication repository was indicative of the work related social network of the developers, and the CTO suggested the bug tracker. This was further reiterated by the different developers and support staff that we interviewed. We also cross checked the social network with the data about the social network from the interviews (when they answered the question "Who they talk to and how often?").

Code Ownership: We relied on the CVS log files, which has been used by others [36], to calculate code ownership. Also the company policy was for all developers to promptly check-in (into the CVS) the code they had modified.

Information centrality: As discussed earlier, we verified with the CTO and developers that the discussions in the bug tracker were indeed representative of the developer-developer social network.

In identification of both the Conway's Law and the Project Coordination STSC, the Social Network of the employees is considered. The research presented in this paper, as in the previous work on identifying the Conway's Law STSC [31, 34] quantitatively, have ignored the content and semantics of the messages. What is important to be considered, is whether the communication between the employees really resolves the problems in the dependencies among the software modules or components. Thus, the content and the semantics of the communication messages between employees have to be considered. Furthermore, the Communication Richness [79] of the messages could be taken into account. Communication Richness as defined by Ngwenyama and Lee (1997) [79] "not only understanding what the speaker or writer means, but the action type associated with the action type enacted by the speaker or writer. The results of the tests enable the listener or reader to identify and analyse distorted communications. By distorted communication, we mean communicative acts that are false, incomplete, insincere or unwarranted." ([79], p152). Thus, one may consider the Communication Richness in each of the messages sent between the employees of an organization while computing the Social Network of the organization.

**Internal Validity**
Internal validity deals with cause and effect relationships. The threat to internal validity is whether "the observed effects could have been caused by or correlated with a set of un-hypothesised and/or unmeasured variables" [80]. This paper claims that the use of TESNA method and tool makes it easier for a Project Manager to identify STSCs. So, the independent variable here is the existence or non-existence of our method and tool, while the dependent variable is the possibility of identifying STSCs.

There are two main threats to the internal validity:
- Selection Bias: That the case study selected made it easier to identify STSCs
- History: The experience of Project Managers in dealing with similar cases made the identification of STSCs easier.

In the first case of selection bias, there is a possibility that the relatively small size of eMaxx made it easier to detect STSCs, but we think that with data clustering and aggregation it is possible to detect these STSCs in much larger case studies.

We think that, except for the Conway's Law STSC, the previous experience of the managers was not influential in identifying the STSCs. When the results were shown to the support staff, developers and project managers in the final presentation session, most were initially surprised by the Code Ownership and Project Coordination STSC. On explaining the results, the employees did agree with the findings. While in the case of Conway's Law STSC, the lack of communication among the different teams (Front Office, Application Server and BPEL Engine) was quite evident and was even bothering some of the project managers, as is evident from their interviews. However, the magnitude of the Conway's Law STSC still came as a surprise to the managers, when the data in Figure 3 were shown to them during the workshop presentation.

**External Validity**

External validity refers to how well the results of the case study can be generalised beyond the study data. Lee and Baskerville [81] provide a framework for the different types of generalizability. They provide suggest four different types of generalizability:
- Type EE: Generalising from data to description
- Type ET: Generalising from description to theory
- Type TE: Generalising from theory to description
- Type TT: Generalising from concepts to theory

The research presented in this paper falls into the category of type ET form of Generalizability. Lee and Baskerville [81] describe two ways of generalizability involved in a type ET generalizability: generalising from empirical data to theory and generalising the resulting theory to other domains and samples.

In this research we showed how we developed our method and tool from empirical data. On performing the case study and from the resulting empirical data we have come to develop a theory that the detection of STSCs is easier with the method and tool that TESNA provides. This theory is a first step to a type ET generalizability [81]. Regarding the second step of generalising the resulting theory beyond the sample, Lee and Baskerville state that such generalizability is not feasible. In other words, they state that the theory resulting from empirical data from a case study cannot be generalized beyond the particular case study [81]. In this context, it is interesting to see if the theory is indeed valid in large (commercial), and in globally distributed settings.

**Reliability**

In the case of Reliability, we have followed a case study protocol as indicated by Yin [75], and have documented the case study. Hence the measures and analysis for the three patterns are repeatable in other settings [75].

**8. CONCLUSION**

An increasing number of Socio-Technical Patterns are becoming available based on experiences and expert opinions. These patterns are potentially useful for managing systems development, but it is difficult and labour intensive for the project manager to select appropriate patterns and keep track of their potential occurrences. Identifying STSCs can prove particularly difficult when multiple people are responsible for various tasks and when the task assignments keep changing in a dynamic and iterative software development environment. Though as team leader David suggested, the detection of a STSC does not necessarily mean that a real STSC exists. Rather, the presence of a STSC can be considered as one more indication of a potential problem in the Socio-Technical structure of the organization. Further investigation, e.g. looking into other archives (like e-mail and chat archives), as done in Amrit and van Hillegersberg [3], can bring more clarity to the results. Though one would expect to find STSCs in large software projects, we were gratified to find the presence of STSCs even in medium sized companies like eMaxx, thus confirming the usefulness of tools such as TESNA even in smaller organizations. It is our conjecture that with increase in size of the software organizations, and an increase in the size and complexity of software, tools like TESNA will provide even greater value.

We are currently conducting additional case studies in order to detect technical dependencies at different levels, for example at the code level, at the level of the architecture and at the level of work flow. Through the investigation of different dependencies we can gain insights into different possible STSCs. We have found that dependencies due to the code structure are more applicable to larger software development organizations.

We are also conducting case studies to study the presence of STSCs in middle and large software development organizations and inter-organisational development in globally distributed settings. We are investigating open source software development [9, 82], to see the differences between corporate STSCs and Open Source STSCs. We are also in the process of using this technique to validate new and existing Socio-Technical patterns. Future research could also focus on different predictors of STSCs rather than only studying the outcomes, as we have done in this research.

A suite or repository of Socio-Technical patterns similar to the Portland Pattern Repository [83] would be useful for future research and for an elaborate testing of the TESNA method and tool.

**Appendix A**

To represent the people and the software in an understandable way we cluster the software into clusters according to the class level dependencies [68] and display who is working at which cluster for the particular time period of the data.

The algorithm we use is as follows:

**Algorithm 1** Dependency Based Clustering Algorithm
**Input:** n software modules, and their $n*n$ DSM, Number of Clusters k
**Output:** k clusters $\{C_1, \ldots, C_K\}$
1: Identify vertical buses
2: Calculate Initial Clustered Cost
3: **repeat**
4:   Select random software module m
5:   Accept bids for m from the clusters
6:   Determine the best bid
7:   If bid is acceptable; modify the clusters
8:   Determine if clusters are stable
9: **until** Clusters are stable
10: Output the Clusters

*Algorithm 1: The algorithm used for clustering the Software Module DSM (adapted from[67])*

In the above algorithm the vertical buses are those elements in the *SM* whose "vertical dependencies" (ones in the vertical columns of the *SM* matrix) to other elements is more than a specific threshold [67]. These elements are important, as they are common functions called by other modules [67]. Once these vertical buses are identified a *DependencyCost* is assigned to each module, element of *SM*. This *DependencyCost* is assigned as follows:

$$DependencyCost(i \rightarrow j \mid j \text{ is a vertical bus}) = d_{ij}$$
$$DependencyCost(i \rightarrow j \mid j \text{ is a vertical bus}) = d_{ij} * n^\lambda$$
$$DependencyCost(i \rightarrow j \mid j \text{ is a vertical bus}) = d_{ij} * N^\lambda$$

*Equation 2: Calculation of the Dependency Cost (taken from [67])*

Where $d_{ij}$ is a binary variable indicating dependency between *i* and *j* (so in our case it is $SM(i,j) + SM(j,i)$), *n* is the size of the cluster when *i* and *j* located within the cluster and *N* is the size of the *SM* matrix (when *i* and *j* are not located in the same cluster). $\lambda$ is a user defined parameter and is found by trial and error (depending on the variation of the results) to be optimum at 2. Adding an element to a cluster increases the cost of other dependencies in the cluster (as the size of the cluster increases), hence an element is only added to a cluster when the reduction in the sum of *DependencyCosts* with the element exceeds the added costs borne by other dependencies[67].

Now the summation of the *DependencyCosts* of all the elements of *SM* gives us the *ClusteredCost* of the matrix for the particular iteration. Hence the *ClusteredCost* can be expressed as:

$$CC(i) = \sum_{j=1}^{n} (SM(i,j) + SM(j,i)) \times size(i,j)^2$$

*Equation 3: Calculation of Clustered Cost (adapted from [68] and [67])*

In Equation 3 *CC(i)* represents the *Clustered Cost* for the element $SM(i,j)$

**Appendix B**
Here we calculate the Core Periphery Distance Metric (CPDM)

```
Algorithm 2 Calculating Core Periphery Distance Metric (CPDM)
Input: Cluster Dependency Structure Matrix (Cluster DSM) that represent the dependencies
between software module clusters
People Cluster Matrix that represent the people working on the clusters,
Number of Clusters k  // We take k = 9
Output: CPDM Metric for Each Developer
  Identify the Core and the Periphery of the Cluster Dependency Matrix
  Assign numbers 0-k for the clusters with 0 being most Core
  Rearrange the columns of the People Cluster Matrix according to the descending
  Core-Periphery Column order of the Cluster Dependency Matrix
  for Each of the developers in the People Cluster Matrix do
    Calculate CPDM = k - (Distance from Core Cluster)
    if Developer has modified source code files in more than one Cluster then
      Calculate Average CPDM
    end if
    Print CPDM
  end for
```

*Algorithm 2: The Core Periphery Distance Metric (CPDM) algorithm*

In the above algorithm, in order to identify the core and the periphery of the *Cluster Dependency Matrix* we realize that the core-ness of a particular cluster depends not only on the size of the cluster but also the dependencies of the particular cluster with other clusters. We hence multiply the *Cluster Dependency Matrix* with the *Cluster Size Matrix* (the matrix with the sizes of the corresponding clusters). The resulting matrix gives us an indication of the core and the periphery clusters with the larger entries being more core than the smaller entries. So if we arrange the columns of this matrix in the descending order we would have the clusters in the descending order of core-ness.

Also in Algorithm 2, the *Distance from the Core Cluster* is given by Equation 5.

$$DCC = \frac{\sum_{j=1}^{k}(d_{ij} * j)}{k}$$

*Equation 4: Calculation of the Distance from the Core Cluster*

In the above Equation 4 $d_{ij}$ represents the *(i,j)* element of the *People Cluster* Matrix, while *k* are the number of columns (9 in our case). So the closer the *CPDM* is to k the more number of clusters in the core the developer has modified

**Appendix C**

Here we describe the demographic data of the interviewees

TABLE 2: THE DEMOGRAPHIC INFORMATION OF THE INTERVIEWEES

| Name | Gender | Role | Team | # Years of Work Experience | # Years of Work Experience for eMaxx |
|---|---|---|---|---|---|
| David | M | CTO, Developer | Application Server | 15 | 10 |
| Ethan | M | Developer | Application Server | 10 | 10 |
| Mason | M | Developer | Database Adaptors | 5 | 5 |
| Thomas | M | Developer | Database Adaptors | 4 | 4 |
| Ryan | M | Developer | Front Office | 5 | 5 |
| Kyle | M | Developer | Front Office | 2 | 2 |
| Oliver | M | Support | Support | 5 | 3 |
| Faron | M | Developer | BPEL Engine | 4 | 4 |
| Lin | F | Manager | Manager | 3 | 3 |
| Max | M | Intern | Intern | 0.5 | 0.5 |